# Large Language Models as Partners in Student Essay Evaluation

Toru Ishida, Tongxi Liu, Hailong Wang, and William K. Cheung

*Abstract*—As the importance of comprehensive evaluation in workshop courses increases, there is a growing demand for efficient and fair assessment methods that reduce the workload for faculty members. This paper presents an evaluation conducted with Large Language Models (LLMs) using actual student essays in three scenarios: 1) without providing guidance such as rubrics, 2) with pre-specified rubrics, and 3) through pairwise comparison of essays. Quantitative analysis of the results revealed a strong correlation between LLM and faculty member assessments in the pairwise comparison scenario with pre-specified rubrics, although concerns about the quality and stability of evaluations remained. Therefore, we conducted a qualitative analysis of LLM assessment comments, showing that: 1) LLMs can match the assessment capabilities of faculty members, 2) variations in LLM assessments should be interpreted as diversity rather than confusion, and 3) assessments by humans and LLMs can differ and complement each other. In conclusion, this paper suggests that LLMs should not be seen merely as assistants to faculty members but as partners in evaluation committees and outlines directions for further research.

## I. INTRODUCTION

To foster creativity, the use of workshops in education is expanding. Workshop courses are conducted as training in collaboration and project management, or as part of active learning. Student assessments utilize a rubric, which breaks down the learning objectives of the workshop [1]. Due to the non-mechanical nature of these assessments, the burden on faculty members is significant. Faculty members can provide high-quality assessments due to their cognitive flexibility but are susceptible to subjectivity and bias [2]. Therefore, multiple faculty members must conduct assessments and deliberate together, making the evaluation process further time-consuming [3].

Recently, applications of Large Language Models (LLMs) have been reported across various research fields, including gaming [4], education [5], journalism [6], healthcare [7][8], drug discovery [9], arts [10], geography [11] and humanities [12]. A common use is guiding humans in information retrieval [6][11][12]. However, there is a perception that LLMs are still inferior to experts. Therefore, in fields where errors are not tolerated, such as healthcare, LLMs act as assistants to experts [7][8]. Efforts to audit LLMs before application are also underway [13]. However, in areas like gaming, where LLMs can surpass human abilities, they take on the role of human partners, and their judgments are directly adopted [4]. So, how should faculty members and LLMs collaborate in the educational context?

To understand the relationship between education and LLMs, this paper explores LLM assessments of student reflective essays. Even before the advent of LLMs, the automatic evaluation of student essays had been studied and applied in writing skill training. For instance, practices such as analyzing linguistic features used in essays and comparing them with model essays have been established. However, in the workshop courses targeted in this paper, it is necessary to delve into the content of essays to evaluate innovative ideas and creative argumentation. While LLMs are considered capable of high-

Toru Ishida was with Waseda University, Tokyo, Japan. He is now with the Department of Computer Science, Hong Kong Baptist University, Kowloon Tong, Kowloon, Hong Kong (e-mail: toru_ishida@comp.hkbu.edu.hk).

Tongxi Liu is with the Department of Educational Studies, Hong Kong Baptist University, Kowloon Tong, Kowloon, Hong Kong (e-mail: tongxi_liu@hkbu.edu.hk).

Hailong Wang was with Waseda University, Tokyo, Japan. He is now with the Department of Civil Engineering, The University of Tokyo, Tokyo, Japan (e-mail: h.wang@civil.t.u-tokyo.ac.jp).

William K. Cheung is with the Department of Computer Science, Hong Kong Baptist University, Kowloon Tong, Kowloon, Hong Kong (e-mail: william@comp.hkbu.edu.hk).

quality assessments using their background knowledge, there are opinions that evaluations cannot be entrusted to them due to their unstable nature and black-box functionality.

To answer these research questions and understand the capabilities of LLMs, we conduct the following two analyses using reflection essays from actual workshop courses:

1) **Quantitative analysis of LLM evaluations**

   We implement three methods of essay evaluation by LLMs: i) evaluation without guidance, ii) evaluation with specified rubrics, and iii) evaluation by pairwise comparison, to measure their assessment capabilities. Additionally, we measure the correlation between faculty members' assessments and those by LLMs.

2) **Qualitative analysis of LLM evaluations**

   While the quantitative analysis indicates high assessment capabilities of LLMs, it leaves concerns about consistency and stability. Therefore, we analyze the content of LLM assessment to understand i) LLMs' rubric-generating behavior, ii) the causes of instability in LLM assessments, and iii) differences between faculty members' assessments and those by LLMs.

These analyses show that LLMs can match faculty members' assessment capabilities, highlighting further research challenges in scenarios where LLMs act as partners to faculty.

## II. HISTORY OF ESSAY EVALUATION

This paper is a collaborative effort between the fields of education and computer science. This section provides an overview of previous research on essay evaluation in education as background to this paper.

In the field of education, various theories have been explored regarding the evaluation of workshop essays. For instance, understanding the content of workshop themes is assessed based on lower-level skills in Bloom's Taxonomy, such as the ability to understand facts and concepts [14]. Based on constructivist theories [15][16], students are evaluated on their ability to construct knowledge from their experiences. Contributions to the team are assessed based on higher-level skills in Bloom's Taxonomy, evaluating how information is analyzed, evaluated, and new ideas are generated. The depth of introspection is evaluated based on Experiential Learning Theory [17] and the theory of the Reflective Practitioner [18], focusing on extracting learning from experiences and ap-plying it to future actions. Additionally, how students perceive and express their growth is evaluated based on Self-Efficacy Theory [19].

Furthermore, the consistency and alignment of learning objectives, teaching methods, and evaluation rubrics [20], the reproducibility of evaluation outcomes under certain conditions [21], and the fairness and unbiasedness of evaluations for all students [22] are deemed crucial.

Essay scoring by human evaluators has been a long-standing practice [23]. The human scoring approach is capable of providing a comprehensive and contextually nuanced assessment of students' essays [24]. This holistic evaluation considers various dimensions of essays, including certain aspects of students' academic writing abilities [25]. Research by Mozer *et al.* [26] demonstrates the value of human interpretive skills in understanding subtleties and context-specific elements.

On the other hand, the history of automated essay evaluation is outlined in [27]. Initially, Project Essay Grader™ evaluated essays based on objective indicators such as sentence length, word choice, and grammatical accuracy. With the advent of the 2000s, artificial intelligence technologies like machine translation and natural language processing began to be incorporated into evaluation. For example, the Intelligent Essay Assessor™ (IEA) recognized word patterns in text and measured the relevance of essays to topics [28]. E-rater, developed by Educational Testing Service, assessed essay quality by analyzing syntactic correctness and vocabulary diversity [29]. CriterionSM is an online writing evaluation service that utilizes E-rater [30]. IntelliMetric™ enhanced evaluation accuracy by incorporating content-related indicators such as idea development and organizational flow [31]. As mentioned above, before the emergence of LLMs, automated essay evaluation was conducted largely based on linguistic features.

III. Case: Workshop Course

The essays analyzed in this paper are from a mandatory workshop course titled "Research Planning and Skills A," offered by the School of Creative Science and Engineering at Waseda University in 2021. The course aims to deepen understanding of technologies and to boost motivation and active participation in group work. Conducted in English, the workshop required participants to write their essays in English as well. This study analyzes essays from 22 participants, covering four themes listed in Table 1.

TABLE I
Four Workshop Themes

| |
|---|
| **1. Kinematic Synthesis (Student IDs 1-6)** <br> Addressing the safety hazard of the gap between trains and platforms, especially for vulnerable populations, the project aims to design a risk-reducing mechanism. |
| **2. All Electric Airplane (Student IDs 7-11)** <br> Aiming to reduce CO2 emissions from airplanes, the project considers an all-electric design as a solution to significantly lower emissions. |
| **3. Radioactive Waste (Student IDs 12-16)** <br> Focusing on high-level radioactive waste from spent nuclear fuel, the project explores disposal methods and strategies to address social concerns. |
| **4. Monster Track (Student IDs 17-22)** <br> Due to oversized loading exceeding bridge capacity and causing collapses, the project seeks solutions to mitigate this issue. |

Students addressed this challenge through seven 3-hour workshops, focusing on ideation, interim reporting, prototyping, and concluding with a reflection essay. The essay was required to include the following sections:

1) **Project Description (300-400 words)**
   The title, goal, and conclusion of the project; the process of the project, including how students applied design thinking methods.
2) **Contribution (300-400 words)**
   The student's role in the project and their contributions to it.
3) **Reflection (200-300 words)**
   Students should write freely, focusing particularly on what they learned from the series of workshops. This includes not only professional knowledge and skills but also insights into how to contribute to and facilitate the workshops.

The submitted essays were evaluated by six faculty members using a five-point scale, and the scores were combined. Each essay could earn up to 30 points, which accounted for 30% of the course grade. Generally, the essays were of good quality, with scores ranging from 22 to 28 points. The scoring rubric was discussed among the faculty members beforehand, but it was not standardized. Some faculty members prioritized students' achievements, while others focused on students' growth. There were those who valued teamwork, and others who appreciated individual contributions. Similarly, some evaluated writing quality, while others resonated with the depth of introspection.

IV. Quantitative Analysis of LLM Evaluations

We present the results of experiments in which LLMs were used to assess student essays by three approaches. These experiments were conducted using ChatGPT-4 from November to December 2023.

1) **LLM Evaluation without Guidance**
   The LLMs are requested to evaluate essays without any criteria. The LLMs autonomously generated rubrics and assessed based on them. Due to the LLMs inherent randomness, different rubrics were generated each time an assessment was made.

2) **LLM Evaluation with Rubrics:**
We instructed the LLMs to evaluate essays using predefined rubrics. The same rubrics were applied in all assessments.

3) **LLM Evaluation by Comparison:**
Using the specified rubrics, essays were compared pairwise, and the LLMs determined which was superior. The same rubrics were applied in all evaluations.

Furthermore, to analyze the LLM evaluations, we conducted separate assessments with human experts.

4) **Human Evaluation:**
Multiple faculty members were asked to evaluate essays using the specified rubrics. The same rubrics were applied in all evaluations.

*A. Human Evaluation*

A controlled evaluation was conducted as a benchmark by four faculty members (authors of this paper), utilizing the specified rubrics and their performance levels as shown in Tables 2 and 3. Each evaluation criterion generally corresponds to individual paragraphs of an essay, but not exactly, to encourage reviewers to assess each criterion while considering the essay as a whole.

TABLE 2
SPECIFIED RUBRICS

| |
|---|
| 1. Technical Knowledge and Application (10 Points) |
| • Understanding of Concepts: The student's grasp of technological and theoretical concepts relevant to the project. |
| • Practical Application: Effectiveness in applying technical knowledge in practical situations. |
| • Innovation and Problem Solving: Creativity and innovation in addressing challenges and proposing solutions. |
| 2. Teamwork and Collaborative Skills (10 Points) |
| • Individual Role and Contribution: The student's clarity in defining and fulfilling their role, contributing to the project. |
| • Team Interaction and Communication: Ability to communicate and collaborate effectively within the team. |
| • Peer Engagement: Participation in peer learning, support to team members, and contribution to team dynamics. |
| 3. Reflective Learning and Personal Growth (10 Points) |
| • Self-Reflection and Insights: Depth of self-reflection on learning and development throughout the project. |
| • Design Thinking and Process: Application of design methods and management of the project process. |
| • Skill and Attitude Development: Growth in soft skills, such as critical thinking, adaptability, and communication. |

TABLE 3
PERFORMANCE LEVELS

| |
|---|
| **Outstanding (10 Points)** |
| Their work demonstrates exceptional understanding, creativity, teamwork, and introspection. They exhibit mastery in applying theoretical knowledge to practical scenarios and show remarkable initiative. |
| **Highly Competent (8 Points)** |
| Their work reflects a thorough understanding and application of concepts, effective collaboration, and thoughtful reflection. They can integrate knowledge with practical skills and show proficiency in problem-solving. |
| **Competent (6 Points):** |
| Their work shows adequacy in understanding and applying concepts, collaborating with the team, and reflective thinking. |
| **Needs Improvement (4 Points)** |
| There is a noticeable lack of depth or understanding in the technical application, collaboration skills, or reflective thinking. |

The faculty members' evaluation was conducted in two phases. Initially, six essays were selected for a sample evaluation, with each faculty member independently assessing the essays. Despite using common rubrics, significant variance in evaluations was observed for some essays. When a reason exists for the significant variance in evaluation scores, averaging them becomes meaningless. Consequently, a meeting was held among the faculty members to identify the reasons for the variance in evaluations, and guidelines for scoring were established as follows.

1) **Expertise of the Evaluators**

   When evaluators possess expertise in the theme of the workshop, they tend to narrow down the technology items for evaluation. For instance, mechanical engineering experts consider CAD tools merely as means, thereby rating essays that detail mastery of these tools lower, whereas non-experts view mastery of CAD tools as within the scope of the workshop's objectives. Furthermore, experts are likely to be stringent in evaluating the novelty of proposals. For example, in a workshop on Electric Aircraft, students might feel their analysis and proposals are groundbreaking, yet to experts, these may seem like replications of existing ideas. To avoid discrepancies in outcomes based on evaluator expertise, it is essential not to overly restrict the technology focus but to adopt a broad perspective. Instead of assessing the novelty of the proposal directly, evaluate whether the process that led to the proposal could have generated novelty.

2) **Preconceptions of the Evaluators**

   Evaluations are often determined by the initial impression received from an essay. This impression usually stems from the evaluator's preferences, obstructing objective assessment. For instance, evaluators who dislike catchy slogans or excessive expressions tend to dismiss essays employing such language as lacking substance. Furthermore, when a faculty member has directly supervised a workshop, the student's daily activities may dominate the essay's assessment. To avoid such biases, it is crucial to ensure that likes or dislikes towards the impression an essay conveys do not hinder the objective reading of its details. Additionally, there is a need for awareness that the evaluation of an essay should be based solely on the essay itself.

3) **Understanding of Rubrics**

   Rubrics are divided into several criteria, each scored independently. However, during evaluation, the scores of different criteria can influence one another. In our case, 'Technical Knowledge and Application' assesses a student's technical proficiency, whereas 'Reflective Learning and Personal Growth' evaluates their development. Confusing these can lead to overly favoring students who have achieved high proficiency, either initially or through the course, neglecting the workshop's aim to foster growth. To avoid this, faculty members must evaluate with a strong awareness that each rubric item is independent.

Following the guidelines established above, faculty members independently evaluated essays from 22 students and collectively determined the scores through discussion. The results are presented in Figure 1. As in the upper-left, the range of evaluation scores varied from 18.33 to 26.33. With a passing score of 18 and scores above 24 qualifying for an A, there were no failures, and seven students received grades of A or higher. The lower represents each student's evaluation scores (maximum, minimum, average), indicating that the range of scores remained stable due to the established scoring guidelines. The upper-right shows the correlation between the evaluations conducted in a real workshop course and our Human Evaluation, with a high correlation coefficient of 0.798.

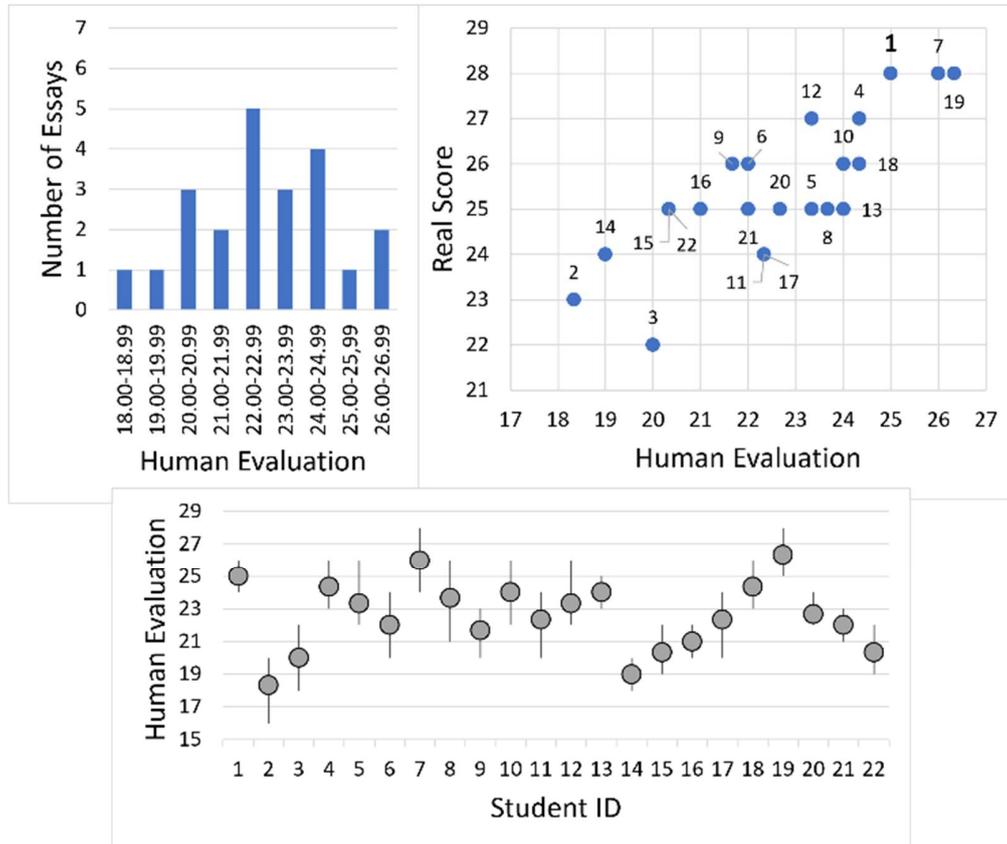

**Fig. 1.** Human Evaluation (Upper-left: Distribution of Human Evaluations. Upper-right: Correlation with Real Score. Lower: Evaluation of Each Essay)

*B. LLM Evaluation without Guidance*

The LLM Evaluation without Guidance method requests that LLMs evaluate student essays without providing any rubrics. The prompt given to the LLMs consists of three parts:

1. **Workshop description:** This is an explanation of the content of the workshop the students attended and the structure of the essays they were assigned, generally around 500 words.
2. **Student essays:** Essays comprising a Project description, Contribution, and Reflection, ranging from 800 to 1100 words.
3. **Instructions for the LLMs:** The evaluation procedure is outlined in approximately 50 words as follows:
   • Generate rubrics based on the workshop description.
   • Explain the theoretical background behind the creation of those rubrics.
   • Evaluate the student's essay out of 30 points using the created rubrics.
   • Explain the reasoning behind the assigned score.

Due to the inherent randomness in LLMs, the generated rubrics vary each time. While it is possible to minimize randomness, it would mean fixing the rubrics randomly. Thus, we accept randomness, and the average score from different rubrics is used for evaluation.

The LLMs evaluated 22 essays six times each, calculating an average score. The results, shown in Figure 2, reveal that scores ranged narrowly from 20.17 to 23.33, as indicated in the upper-left. There were no failures, but also no scores above an A. This is because each student's score is obtained by averaging six different rubrics. The lower part shows the variation in individual student scores before averaging, which is wider than in Human Evaluation due to the different rubrics used. There is a contradiction between allowing diverse evaluation criteria and averaging them. These findings suggest that when different rubrics are accepted, it is crucial not just to average the scores but to engage in a thorough discussion to converge on an evaluation score. The upper-right shows the correlation between LLM and faculty member evaluations, with a correlation coefficient of 0.611, indicating a not sufficiently high correlation.

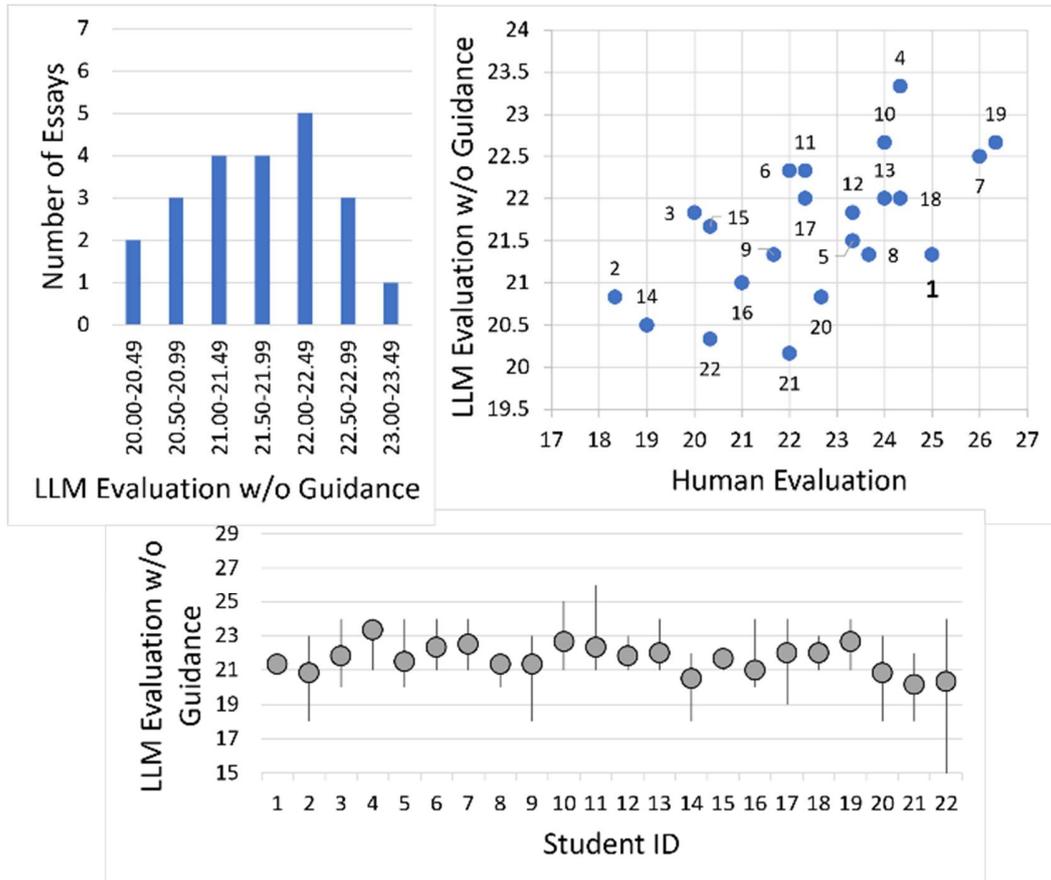

**Fig. 2.** LLM Evaluation without Guidance (Upper-left: Distribution of Evaluations. Upper-right: Correlation with Human Evaluation. Lower: Evaluation of Each Essay)

*C. LLM Evaluation with Rubrics*

The evaluation method, LLM Evaluation with Rubrics, involves requesting LLMs to evaluate student essays using specified rubrics as the evaluation criteria. The prompt consists of the following four parts:
1. **Workshop description:** As previously mentioned.
2. **Student essays:** As previously mentioned.
3. **Rubrics:** The same ones used in Human Evaluation are included in the prompt.
4. **Instructions for the LLMs:** The evaluation procedure is outlined in approximately 50 words as follows:
    • Evaluate the student's essay out of 30 points using the specified rubrics.
    • Explain the rationale behind the assigned score.

Although the rubrics are specified, the evaluation results are not uniform due to the inherent randomness of LLMs. Therefore, for 22 essays, six evaluations were conducted for each, and their average scores were calculated. The results are shown in Figure 3. As indicated in the upper-left, scores ranged widely from 17.00 to 24.33. There were two failures and only one score above an A, indicating stricter outcomes compared to Human Evaluation.

The lower part represents the evaluation scores for each student. The scores before averaging displayed a wider range than those in Human Evaluation. Although rubrics are specified, some essays receive varied scores, undermining the purpose of averaging them. The cause of this variation in scores will be clarified through the qualitative analysis in the next section. The upper right section shows the correlation between LLM and Human Evaluations. The correlation coefficient is 0.657, indicating a stronger correlation than LLM Evaluation without Guidance; however, it is still not sufficient.

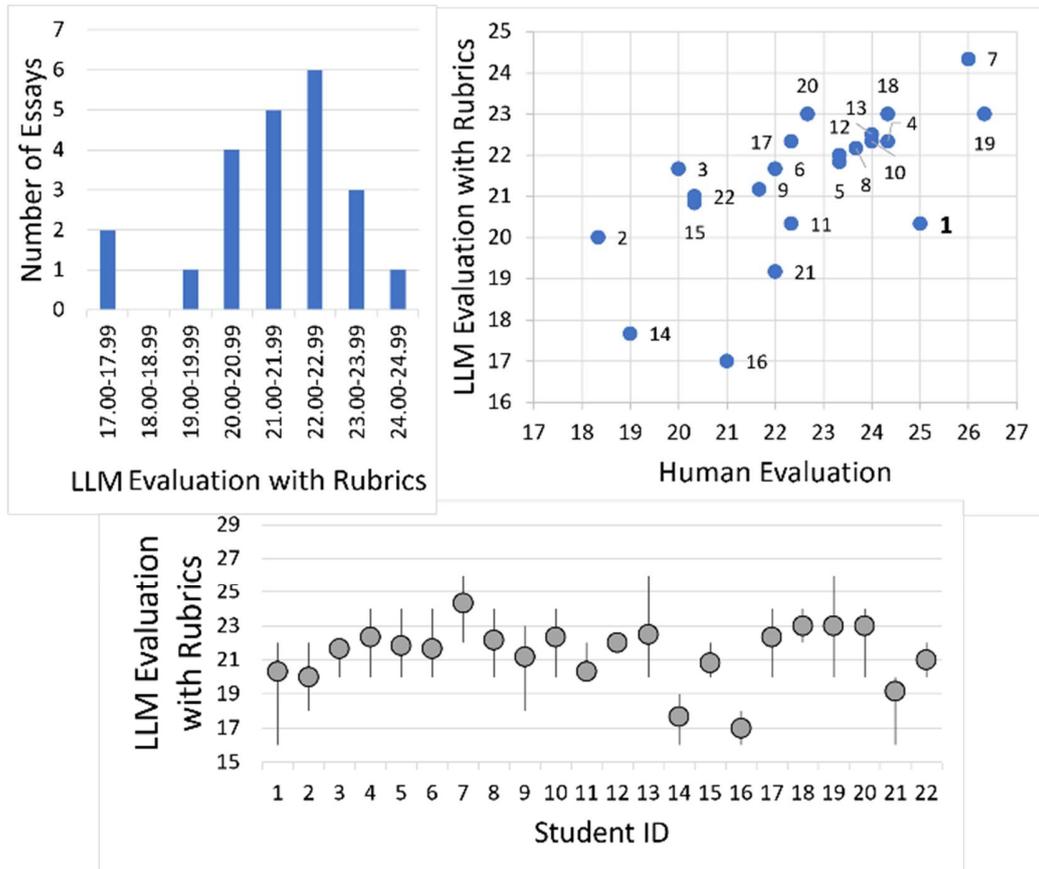

**Fig. 3.** LLM Evaluation with Rubrics (Upper-left: Distribution of Evaluations. Upper-right: Correlation with Human Evaluation. Lower: Evaluation of Each Essay)

*D. LLM Evaluation by Comparison*

Despite the use of standardized rubrics in LLM evaluations, variability in student assessments by LLMs remains. The reason is that LLMs review essays on an individual basis, ignoring the comparative ranking of essays. To tackle this issue, a method was introduced for LLMs to compare essays pairwise. The prompt consists of four parts:

1. **Workshop description:** As previously mentioned.
2. **Student essays:** As previously mentioned, but with two essays included.
3. **Rubrics:** The same ones used in Human Evaluation are included in the prompt except Performance Levels.
4. **Instructions for the LLMs:** The evaluation procedure is outlined in approximately 100 words as follows:
   - LLMs compare two essays based on three main criteria: Project Description, Role and Contribution, and Reflection and Insight.
   - If one essay is slightly superior, it receives +1 point, and the other -1 point. If one is clearly superior, it receives +2 points, and the other -2 points. If both are equal, each receives 0 points.

A round-robin tournament involving 22 students' essays, resulting in 231 pairwise evaluations, was conducted using LLM. The scoring was simplified as follows:

1. Sum the points from the comparative evaluations for each essay and normalize this sum within a range of 0 to 1.

2. Map the normalized scores onto a range from 18 to 30 points, assuming there are no failures.

The scores from the pairwise evaluations ranged from 19.14 to 29.29, as indicated in the left of Figure 4. Pairwise comparisons often result in distinct winners and losers. Consequently, when calculating evaluation scores, it is crucial to mitigate the extreme disparities between higher- and lower-ranked students. The right of the figure illustrates the correlation between LLM and Human Evaluations, with a correlation coefficient of 0.716. This demonstrates a stronger correlation compared to LLM Evaluation with Rubrics. However, the comments from the LLMs in this evaluation, which only discuss pairwise comparisons, are not appropriate for providing feedback to students.

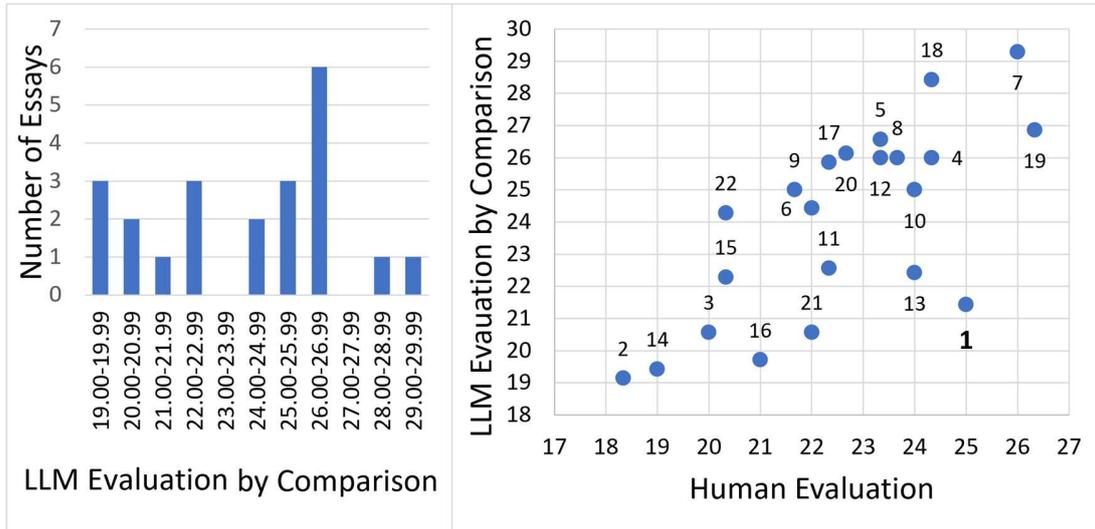

**Fig. 4.** LLM Evaluation by Comparison (Left: Distribution of Evaluations. Right: Correlation with Human Evaluation.)

Pairwise comparison results differed from the final evaluation ranking in 16.9% (39 instances) of all comparisons. If one were to conduct comparisons of evaluation scores in each of the six instances of LLM Evaluation with Rubrics, the proportion of pairs differing from the final evaluation ranking would average 35.5% (82 instances). Although it is not straightforward to conclude due to the differences in evaluation methods, these numbers suggest that pairwise comparisons yield more stable evaluations.

## V. QUALITATIVE ANALYSIS OF LLM EVALUATIONS

We conducted a qualitative analysis of three interesting questions identified in the quantitative analysis.

### A. How Rubrics Created by LLMs Vary

In 132 rubrics generated by the LLM Evaluation without Guidance, two distinct patterns have been identified. Pattern 1 involves conducting evaluations for each section of an essay, as shown in Table 4. Utilizing such rubrics often results in higher scores due to the direct correspondence between the sections of the essay and the evaluation criteria. However, this approach makes it difficult to assess the overall understanding or depth of learning presented throughout the essay.

TABLE 4
LLM GENERATED RUBRICS (PATTERN 1)

| |
|---|
| **1. Project Description (10 Points)** |
| • Clarity and Completeness: Clear explanation of the project's title, goals, conclusions, and the project process. |
| • Insight and Depth: Depth of understanding and insight into the project's challenges and solutions. |
| **2. Contribution (10 Points)** |
| • Role Clarity: Explanation of the student's role in the project. |
| • Impact of Contribution: Detailing specific contributions and their impact on the project. |
| **3. Reflection (10 Points)**<br>• Self-awareness and Learning: Insights into personal growth, learning from the workshop, and group dynamics.<br>• Application and Future Implications: How the learning applies to future endeavors and understanding of its implications. |

Pattern 2 aims to evaluate the essay from various perspectives across its entirety, which tends to result in lower evaluation scores due to the lack of correspondence between the essay's structure and the evaluation criteria. An example is provided in Table 5. In this case, the evaluation of each essay section is aggregated into the first evaluation criterion. The second criterion assesses the essay's clarity and structure based on writing pedagogy. The third criterion is based on reflective practice theories.

TABLE 5
LLM GENERATED RUBRICS (PATTERN 2)

| |
|---|
| **1. Content (10 Points)**<br>• Project Description: Clarity and completeness of the project description, including the title, goals, process, and conclusion.<br>• Role & Contribution: Clear articulation of the student's role and contributions to the project.<br>• Reflection: Depth and insightfulness of the reflection, including lessons learned and personal growth. |
| **2. Clarity & Organization (10 Points)**<br>• Structure: Logical organization and flow of the essay.<br>• Language & Grammar: Clear, concise, and correct writing.<br>• Engagement & Style: Ability to engage the reader and effectively convey information. |
| **3. Analysis & Critical Thinking (10 Points)**<br>• Insight & Depth: Understanding of the project and its implications.<br>• Problem-Solving & Innovation: Evidence of creative thinking and problem-solving skills.<br>• Application of Learned Concepts: Effective application of design thinking methods and other learned concepts. |

The rubrics generated by LLMs are diverse, but it has become clear that each rubric is generated under a clear policy.

*B. How Comparison Stabilizes Evaluations*

To understand how comparative evaluation contributes to the stability of assessments, an analysis was conducted on Student Essay #13 and #19. In the LLM Evaluation with Rubrics, both essays underwent six evaluations, with scores ranging broadly from 20 to 26 points; the variances in the scores were significant, but their average scores were nearly identical.

Table 6 presents the evaluation comments for #13 and #19, citing instances from the assessment criterion 'Reflective Learning and Personal Growth,' including both a positive evaluation (26 points) and a negative evaluation (20 points). Although the understanding of the essays was almost the same, the negative evaluation demanded a higher level of description (italicized). As mentioned previously, Performance Levels within the rubric are divided into four levels, but the assignment to a level depends on the evaluator's standard. A human evaluator would likely conduct comparative evaluations after reading all essays, thus adjusting the scores accordingly. However, with LLMs, essays are assessed individually, resulting in wide variability in the LLM evaluation scores, even though the rubric is specified.

11TABLE 6
POSITIVE AND NEGATIVE EVALUATION COMMENTS

| |
|---|
| **Self-Reflection and Insights (Comments for #13)** |
| Positive: The student offers thoughtful insights into their learning process, acknowledging the importance of teamwork and the complexities of designing sensitive questions. |
| Negative: The student reflects on the importance of teamwork and understanding the sensitivity of their project's topic. *However, deeper personal insights or transformative learning experiences are not extensively discussed.* |
| **Skill and Attitude Development (Comments for #19)** |
| Positive: The essay reflects growth in professional skills like research and creativity, and soft skills such as communication and confidence in group settings. |
| Negative: The student exhibits growth in professional and soft skills, such as critical thinking, creativity, and communication, *though the depth of development could be further elaborated.* |

However, when conducting LLM Evaluation by Comparison, comparing #13 and #19 six times, the results consistently favored #19 in all comparisons. This result is compatible with Human Evaluation. Table 7 shows two comparison examples regarding 'Reflective Learning and Personal Growth.' The reasons for determining superiority or inferiority (italicized) were consistent, resulting in stable evaluations.

TABLE 7
EVALUATION COMMENTS BY COMPARISON

| |
|---|
| Both students showed good reflective learning, but *#19's essay demonstrates a broader spectrum of personal growth, touching on various aspects such as confidence, creativity, and interdisciplinary learning.* |
| Both essays show strong reflective learning and personal growth, with each student demonstrating insight into their development process. However, *#19's reflection on interdisciplinary learning and the practical application of skills gives a slight edge in demonstrating broader personal growth.* |

*C. How Human and LLM Evaluations Differ*

We examined the differences in evaluations of Student Essay #1, presented in Table 8, by faculty members and LLMs. This investigation was initiated because the evaluations by faculty members significantly exceeded those by LLMs for Student Essay #1, as illustrated in the correlation graphs in Figures 2, 3, and 4. Furthermore, the essay also received high marks in the actual workshop course.

TABLE 8
SUMMARY OF STUDENT ESSAY #1

| |
|---|
| Technical Knowledge and Application: The project focused on creating mechanisms incorporating pivots, gears, and hydraulics. I developed four designs; though not selected for presentation, demonstrate an engagement with technical problem-solving. |
| Teamwork and Collaborative Skills: I helped refine communication within the team and offered support for final presentation preparation, enhancing team dynamics and project progression. |
| Reflective Learning and Personal Growth: Overcoming public speaking anxiety and improving presentation and critical thinking skills marked my personal growth. |

Comments from both faculty members and LLMs on Essay #1 are shown in Table 9. LLMs focus on the fact that the four ideas proposed by the student were not adopted, whereas faculty members recognize the student's excellent technical skills through these four ideas. LLMs express concerns that the student's activities were primarily individual research. Faculty members share this view but also appreciate the student's clear awareness of their role, interpreting the limited group activities as a reflection of the student's individuality. While LLMs note the absence of specific growth, faculty members are empathetic towards the student's recognition of and efforts to address their own shortcomings.

Thus, unlike LLMs, faculty members can identify a student's superior abilities and deep introspection from the text. On the other hand, it can also be said that LLMs evaluate essays more objectively compared to faculty members. Furthermore, it is noteworthy that there are no essays where LLM evaluations significantly outperformed those by faculty members, indicating that the evaluations by LLMs are not overly aggressive. From this, it can be inferred that when faculty members and LLMs collaborate on evaluating essays, faculty members should devote more time to uncovering students with exceptionally exceptional qualities.

TABLE 9
COMMENTS FROM LLMS AND FACULTY MEMBERS

| **Comments from LLMs:** |
| --- |
| - His actual contribution to the final project seems limited. |
| - The essay suggests a more individualistic approach than a collaborative team effort. |
| - The student acknowledges improvement in presentation skills, critical thinking, and expressing ideas, but specific examples or instances of these developments are limited. |
| **Comments from faculty members:** |
| - This student described details of his four designs, giving me an impression that he understands the technical issue well. |
| - This student described his work and suggestions and tried to explain why having these ideas. |
| - This student has well identified his shortcomings in collaboration and has overcome them. |

VI. CONCLUSION

In this paper, we analyze the results of evaluating student essays written in workshop courses using LLMs. Although LLMs have begun to be used for evaluations of writing skills, assessing essays from workshop courses—which demand appraisal of originality and collaboration quality—proves to be exceptionally challenging. By conducting over one thousand evaluations and analyzing the data both quantitatively and qualitatively, we have elucidated the following findings.

Firstly, LLMs possess evaluation capabilities comparable to those of faculty members. As demonstrated in LLM Evaluation without Guidance, LLMs can generate diverse rubrics based on pedagogical knowledge. Additionally, LLMs are capable of evaluating essays according to a specified rubric and generating appropriate comments. While determining the Performance Level requires a relative evaluation against other essays, the simultaneous assessment of all essays is not practical. Therefore, an effective method involves repeatedly comparing pairs of essays to determine the overall ranking, which yields a strong correlation, exceeding 0.7, with faculty evaluations. Consequently, it is appropriate to use rubrics for providing feedback to students, and grades should be determined through comparative evaluation among essays.

Secondly, the randomness observed in evaluations conducted by LLMs is not confusion but diversity with logical consistency. While the outcomes of LLM assessments vary with each execution, examining the results reveals that each has a coherent rationale. It is as though a variety of faculty members within the LLM are emerging in succession to express their opinions. This paper proposes a framework for achieving stable evaluations, including specifying rubrics and conducting pairwise comparisons, to converge the diverse opinions within the LLM. However, it also considers the potential of actively utilizing this diversity. By clarifying the basis of the diverse opinions and synthesizing them, we may realize a holistic evaluation.

Thirdly, there are instances where LLMs cannot highly evaluate essays that faculty members are impressed by. LLMs lack the accumulated teaching experience to recognize a student's unique abilities or potential for future growth. On the other hand, LLM evaluations can mitigate the excessive personal bias from faculty members. The question then arises: How should faculty members and LLMs divide their roles? In settings with a faculty shortage, it is conceivable to combine evaluations from a few faculty members and LLMs and compare the results. LLMs have the capability to participate in essay evaluations not just as assistants but as members of the evaluation committee.

Applying LLMs in educational evaluations presents various challenges that remain to be addressed. The societal acceptance of student evaluations by LLMs needs to be investigated. Additionally, examining the transparency of evaluations is necessary when LLMs participate as faculty partners. Concerns about students using LLMs to generate essays also emerge. Including student learning activities, it is crucial to design a learning process that involves collaboration among faculty members, students, and LLMs.

REFERENCES[1] Heidi Goodrich Andrade. Using Rubrics to Promote Thinking and Learning. Educational Leadership 57(5):13–19, 2000.
[2] Nyoli Valentine, Steven Durning, Ernst Michael Shanahan and Lambert Schuwirth. Fairness in Human Judgement in Assessment: A Hermeneutic Literature Review and Conceptual Framework. Advances in Health Sciences Education, 26, pages 713–738, 2021.
[3] Dadi Ramesh and Suresh Kumar Sanampudi. An Automated Essay Scoring Systems: A Systematic Literature Review. Artificial Intelligence Review, 55(3), pages 2495-2527, 2022.
[4] Meta Fundamental AI Research Diplomacy Team (FAIR) et al. Human-level Play in the Game of Diplomacy by Combining Language Models with Strategic Reasoning. Science, 378(6624):1067-1074, 2022.
[5] Kasra Lekan and Zachary A. Pardos. AI-Augmented Advising: A Comparative Study of ChatGPT-4 and Advisor-based Major Recommendations. In Proceedings of NeurIPS Workshop on Generative AI for Education, the Thirty-seventh Conference on Neural Information Processing Systems, 2023.
[6] Savvas Petridis, Nicholas Diakopoulos, Kevin Crowston, Mark Hansen, Keren Henderson, Stan Jastrzebski, Jeffrey V. Nickerson, Lydia B. Chilton. AngleKindling: Supporting Journalistic Angle Ideation with Large Language Models, In Proceedings of the CHI Conference on Human Factors in Computing Systems, 2023.
[7] Eunkyung Jo, Daniel A. Epstein, Hyun-hoon Jung and Young-Ho Kim. Understanding the Benefits and Challenges of Deploying Conversational AI Leveraging Large Language Models for Public Health Intervention. In Proceedings of the CHI Conference on Human Factors in Computing Systems, 2023.
[8] Karan Singhal, Shekoofeh Azizi, Tao Tu, S. Sara Mahdavi, Jason Wei, Hyung Won Chung, Nathan Scales, Ajay Tanwani, Heather Cole-Lewis, Stephen Pfohl, Perry Payne, Martin Seneviratne, Paul Gamble, Chris Kelly, Abubakr Babiker, Nathanael Schärli, Aakanksha Chowdhery, Philip Mansfield, Dina Demner-Fushman, Blaise Agüera y Arcas, Dale Webster, Greg S. Corrado, Yossi Matias, Katherine Chou, Juraj Gottweis, Nenad Tomasev, Yun Liu, Alvin Rajkomar, Joelle Barral, Christopher Semturs, Alan Karthikesalingam and Vivek Natarajan. Large Lan-guage Models Encode Clinical Knowledge. Nature 620, pages 172–180, 2023.
[9] Tianhao Li, Sandesh Shetty, Advaith Kamath, Ajay Jaiswal, Xiaoqian Jiang, Ying Ding and Yejin Kim. CancerGPT for Few Shot Drug Pair Synergy Prediction Using Large Pretrained Language Models. npj Digit. Med. 7, 40, 2024.
[10] Tyler Angert, Miroslav Suzara, Jenny Han, Christopher Pondoc and Hariharan Subramonyam. Spellburst. A Node-based Interface for Exploratory Creative Coding with Natural Language Prompts. In Proceedings of the 36th Annual ACM Symposium on User Interface Software and Technology. Article 100, pages 1–22, 2023.
[11] Rohin Manvi, Samar Khanna, Gengchen Mai, Marshall Burke, David B. Lobell and Stefano Ermon. GeoLLM: Extracting Geospatial Knowledge from Large Language Models, In Proceedings of the Twelfth International Conference on Learning Representations, 2024.
[12] Maria Antoniak, Melanie Walsh, David Mimno and Matthew Wilkens. BERT for Social Sciences and Humanities. In Proceedings of International Conference for Web and Social Media, 2022.
[13] Charvi Rastogi, Marco Tulio Ribeiro, Nicholas King, Harsha Nori and Saleema Amershi. Supporting Human-AI Collaboration in Auditing LLMs with LLMs. In Proceedings of the 2023 AAAI/ACM Conference on AI, Ethics, and Society, pages 913–926, 2023.
[14] Benjamin S. Bloom, Max D. Engel-hart, E. J. Furst, Walker H. Hill, and David R. Krathwohl. Taxonomy of Educational Objectives: The Classification of Educational Goals. Handbook I: Cognitive Domain. New York: David McKay, 1956.
[15] Jean Piaget. The Construction of Reality in the Child. Ballantine, 1954.
[16] Lev Semenovich Vygotsky and Michael Cole. Mind in Society: Development of Higher Psychological Processes. Harvard University Press, 1978. uctor Data Manual, Motorola Semiconductor Products Inc., Phoenix, AZ, USA, 1989.
[17] David A. Kolb, Richard E. Boyatzis and Charalampos Mainemelis. Experiential Learning The-ory: Previous Research and New Directions. In Perspectives on Thinking, Learning, and Cognitive Styles, pages 227-247. Routledge, 2014.
[18] Donald A. Schön. The Reflective Practitioner: How Professionals Think in Action. Routledge, 2017.
[19] Albert Bandura. Self-Efficacy: Toward a Unifying Theory of Behavioral Change. Psychological Review 84(2):191, 1977.
[20] John Biggs. Enhancing Teaching through Constructive Alignment. Higher Education 32(3):347-364, 1996